\documentclass[a4paper,11pt]{article}
\usepackage{amssymb}
\usepackage{amsfonts}
\usepackage{amsmath}

\input{tcilatex}

\begin{document}

\author{Jia Duojie$^{\thanks{%
E-mail: jduojie@yahoo.com }}$ \\
Institute of Theoretical Physics, College of Physics\\
and Electronic Engineering, Northwest Normal \\
University,\textit{\ Lanzhou 730070, China,}\\
Department of Physics, North Carolina State University, \\
Raleigh, NC, 27695-8202}
\title{Abelian-Higgs Phase of SU(2) QCD and Glueball Energy$^{\thanks{%
This work is supported by National Natural Science Foundation of China (No.
10547009) and Research Backbone Fostering Program of Knowledge and S\&T
Innovation Project of NWNU (KJCXGC 03-41).}}$. }
\maketitle

\begin{abstract}
It is shown that SU(2) QCD admits an dual Abelian-Higgs phase, with a Higgs
vacuum type of type-II superconductor. This is done by using connection
decomposition for the gluon field and the random-direction approximation.
Using bag picture with soft wall, we presented a calculational procedure for
glueball energy based on the recent proof for wall-vortices [Nucl. Phys. B
741(2006)1].

PACS number(s): 12.38.-t, 11.15.Tk, 12.38.Aw

\end{abstract}

\section{Introduction}

Recently, the multi-vortices, of the Abrikosov-Nielsen-Olesen type, are
found to be wall vortices\ for the Abelian-Higgs (AH) model \cite%
{BolognesiNum06}. Such a multi-vortices is a bag object with a wall tension $%
T_{W}$ and thickness that separates an internal region with energy density $%
\Delta \varepsilon $ and an external region with energy density $0$. This
provides a novel mechanism for bag objects formation in field-theoretical
framework.

In our previous work \cite{Jiadj07}, an dual AH model was derived from
Yang-Mills (YM) theory and the dual superconductor vacuum is then
investigated. In this paper, we show that the SU(2) QCD admits an dual
Abelian-Higgs phase, with a Higgs vacuum type of type-II superconductor.
This is done by applying connection decomposition \cite{Duan,Cho80,FN} for
the gluon field and the random-phase approximation for the field in QCD
vacuum state. Based on the bag picture of hadron that bag is made of
wall-vortices a calculation procedure for glueball energy is presented for
SU(2) QCD.

Our study is also inspired by the natural emergence of the partial
"electric-magnetic" duality and a gauge-invariant scalar kernel $Z(\phi )$
in the reformulated YM theory \cite{FN,Langmann} and for the effective
confining model of QCD suggested by 't Hooft \cite{tHooftA03}. In the later, 
$Z(\phi )$ assumes the role of the vacuum medium factor, quite similar to
the dia-chromoelectric constant in the dia-chromoelectric soliton (DCS)
model \cite{TDLee,WIL89}. Now that the bag object can arise in the AH model
as a many-vortices soliton, namely, wall vortices \cite{BolognesiNum06} it
is interesting to investigate the QCD origin of dual AH model, the dual and
relativistic version of the Ginzburg-Landau theory for superconductor.

\section{The duality in SU(2) QCD and hadronic picture}

We begin with the $SU(2)$ YM\ theory reformulated by reparameterization
called connection decomposition (CD) \cite{Duan,Cho80,FN}. The gluon field $%
\vec{A}_{\mu }$ (the arrow denotes the three color indices $a=1,2,3$, along
the generator $\tau ^{a}$) is decomposed into \cite{Duan,Cho80} $\vec{A}%
_{\mu }=A_{\mu }\hat{n}+g^{-1}\partial _{\mu }\hat{n}\times \hat{n}\mathbf{+}%
\vec{b}_{\mu }$, in which $\vec{b}_{\mu }$ can be further decomposed into $%
\vec{b}_{\mu }=g^{-1}[\phi _{1}\partial _{\mu }\hat{n}+\phi _{2}\partial
_{\mu }\hat{n}\times \hat{n}]$ \cite{FN} when one considers only the
transverse degrees of freedom. Here, $A_{\mu }$ is an Abelian potential and $%
\hat{n}$ is an unit iso-vector. As a result, one has the Faddeev-Niemi
decomposition \cite{FN} 
\begin{equation}
\vec{A}_{\mu }=A_{\mu }\hat{n}+\vec{C}_{\mu }+g^{-1}\phi _{1}\partial _{\mu }%
\hat{n}+g^{-1}\phi _{2}\partial _{\mu }\hat{n}\times \hat{n}  \label{deFN}
\end{equation}%
with $\vec{C}_{\mu }:=$ $g^{-1}\partial _{\mu }\hat{n}\times \hat{n}$ the
non-Abelian magnetic potential. Here, we have put (\ref{deFN}) in the form
that $\phi $ is of dimensionless. The Abelian magnetic field $H_{\mu \nu }/g=%
\hat{n}\cdot (\partial _{\mu }\hat{n}\times \partial _{\nu }\hat{n})/g$ can
be defined via explicitly calculating the non-Abelian magnetic field tensor $%
\vec{C}_{\mu \nu }=-g^{-1}H_{\mu \nu }\hat{n}$ corresponding to $\vec{C}%
_{\mu }$. We note that the covariance of $\mathbf{b}_{\mu }$ under the gauge
rotation $U(\alpha \hat{n})=\exp (i\alpha n^{a}\tau ^{a})$ ($n^{a}$ is the $%
a $-component of $\hat{n}$) yields the transformation $\phi \rightarrow \phi
e^{-i\alpha }$ for the complex field $\phi :=\phi _{1}+i\phi _{2}$, showing
that it forms a charged complex scalar. This idea of CD is closely
associated with the Abelian projection \cite{tHooftB455}, and can be
generalized to the spinorial-decomposition case \cite{Jinstant,JDJ}.

With (\ref{deFN}), the YM Lagrangian becomes \cite{Langmann} 
\begin{equation}
\mathfrak{L}^{YM}=-\frac{1}{4}[F_{\mu \nu }-\frac{Z(\phi )}{g}H_{\mu \nu
}]^{2}-\frac{1}{4g^{2}}\{(n_{\mu \nu }-iH_{\mu \nu })(\nabla ^{\mu }\phi
)^{\dag }\nabla ^{\nu }\phi +h.c\},  \label{dual}
\end{equation}%
where $F_{\mu \nu }:=\partial _{\mu }A_{v}-\partial _{v}A_{\mu }$, $Z(\phi
):=1-|\phi |^{2}$ and $n_{\mu \nu }:=\eta _{\mu \nu }(\partial \hat{n}%
)^{2}-\partial _{\mu }\hat{n}\cdot \partial _{\nu }\hat{n}$. $\nabla _{\mu
}\phi :=(\partial _{\mu }-igA_{\mu })\phi $ is the $U(1)$ covariant
derivative induced by the gauge rotation $U(\alpha \hat{n})$. We note that
when $A_{\mu }=0$ theory becomes 
\begin{equation}
\mathfrak{L}^{M}=-\frac{Z(\phi )^{2}}{4g^{2}}H_{\mu \nu }^{2}-\frac{1}{4g^{2}%
}\{(n_{\mu \nu }-iH_{\mu \nu })(\partial ^{\mu }\phi )^{\dag }\partial ^{\nu
}\phi +h.c\},  \label{LM}
\end{equation}%
in which the media-like factor $Z(\phi )$ resembles the dia-electric factor
in the DCS model \cite{TDLee,WIL89} and the gauge-invariant kernel in the
effective model \cite{tHooftA03} accounting for the QCD vacuum effects: $%
Z(\phi \rightarrow 0)=1$ in perturbative (normal) vacuum (say, inside
hadrons) and $Z(\phi \rightarrow \phi _{0})\neq 0$ $\,$in the
non-perturbative (NP) vacuum (say, outside hadrons).

The topological variable $\hat{n}(x)$, which defines the homotopy $\pi
_{2}(V)$ of the relevant region $V$, plays the role of the singular
transformation from the global basis \{$\tau ^{1\sim 3}$\} to the local
basis \{$\hat{n},\partial _{\mu }\hat{n}\mathbf{,}\partial _{\mu }\hat{n}%
\times \hat{n}$\}. This suggests that QCD vacuum can be different
topologically with the perturbative vacuum due to the nontrivial homotopic
class of map $\hat{n}$. The validity of the local basis in region $V$
depends upon the regularity of $\partial _{\mu }\hat{n}$ in $V$ which is
broken at isolated singularities $z_{i}$. Note that in the slowly-varying
limit of $\hat{n}$ (that is, the norm $||\partial _{\mu }\hat{n}||$ is
averagely negligible), the decomposition (\ref{deFN}) ceases to make sense
due to the degeneracy of \{$\hat{n},\partial _{\mu }\hat{n}\mathbf{,}%
\partial _{\mu }\hat{n}\times \hat{n}$\}, and in that case one can use the
commonly-used expression $A^{a}\tau ^{a}$ instead.

In the DCS model \cite{TDLee,WIL89} for hadron, theory admits two vacua: one
is the perturbative vacuum with scalar $\sigma =0$ inside the soliton and
other is the non-perturbative vacuum $\sigma =\sigma _{0}$ outside soliton.
The soliton is the field-theory counterpart of the bag in the bag model \cite%
{TDLee,WIL89}. Comparing with this idea about two vacua, one finds that it
is very suggestive to consider the small-$g$ limit of the dynamics (\ref%
{dual}) by assuming $\langle ||\partial \hat{n}||\rangle \sim O(g)$ and $%
\partial \phi \sim o(g)$ as $g\rightarrow 0$. This yields $\langle ||H_{\mu
\nu }||\rangle /g$ $\rightarrow 0$ $\langle ||n_{\mu \nu }-iH_{\mu \nu
}||\rangle /g^{2}\rightarrow const$. The theory then becomes an Abelian
electrodynamics%
\begin{equation}
\mathfrak{L}^{E}=-\frac{1}{4}F_{\mu \nu }^{2}.  \label{LE}
\end{equation}

Let us consider a bag-like picture of a glueball or a hadron with the two
vacua separated by bag boundary region. We assume the existence of the fixed
point of beta function and $g$ $\rightarrow g_{s}$ monotonously as position $%
x$ going from the bag center $\mathbf{x}=0$ to $|\mathbf{x}|=+\infty $ (see 
\cite{Gribov}). Two limits $g$ $\rightarrow 0$ (the ultraviolet limit) and $g
$ $\rightarrow g_{s}\sim 1$ (the ultraviolet limit) correspond to the
perturbative vacuum inside the bag (or soliton) and the NP vacuum outside,
respectively. The dual structure of QCD in these two limits implies that
asymptotically one can view the model (\ref{LE}) as the chromo-electric
dynamics for the inside of bag while (\ref{LM}) as the chromo-magnetic
dynamics for the outside.

To reconcile the bag picture with the dual superconductor mechanism of the
confinement \cite{Nambu} one need to set the average norm $||\partial \hat{n}%
||=\langle (\partial \hat{n})^{2}\rangle ^{1/2}\rightarrow 0$ as $g$ $%
\rightarrow 0$ and the magnetic field fluctuatation $\langle (H_{\mu \nu
})^{2}\rangle ^{1/2}\propto \langle (\partial \hat{n})^{2}\rangle $ $%
\rightarrow H$ (a constant) increasingly as $|\mathbf{x}|\rightarrow +\infty 
$ since $||\partial \hat{n}||$ measures the density of monopoles which
should tend to vanishing inside bag ($g\approx 0$). This implies, as $|%
\mathbf{x}|$ goes from $0$ to $+\infty $, the monopoles density increases,
say, from $\rho =0$ to $\rho _{0}$, since the sites of singularities in the
magnetic field $\vec{C}_{\mu \nu }=-g^{-1}H_{\mu \nu }\hat{n}$ increase as $%
\hat{n}(x)$ is going to vary dramatically. This agrees qualitatively with
the Abelian projection \cite{tHooftB455} that the QCD vacuum is in the
condensed monopoles system, with the normal vacuum penetrated by the
chromo-electric flux-tubes $F_{\mu \nu }$.

\section{Multi-monopoles in the magnetic vacuum}

We consider qualitative behavior of the monopole density $\rho _{m}(\mathbf{x%
})$. As is known, the magnetic charge density is given by \cite{Duan} 
\begin{eqnarray}
\rho _{ch}(\mathbf{x}) &=&\frac{1}{4\pi }\epsilon ^{ijk}\epsilon
_{abc}\partial _{i}n^{a}\partial _{j}n^{b}\partial _{i}n^{c}  \label{rho} \\
&=&\sum_{i}\frac{w(\mathbf{z}_{i})}{g}\delta ^{3}(\mathbf{x}-\mathbf{z}_{i})
\notag
\end{eqnarray}%
where $w(\mathbf{z}_{i})$ stands for the winding number of the map $\hat{n}%
(x)$ at the singularity (monopole) $\mathbf{z}_{i}$. The total magnetic
charge $G_{m}=\int_{V_{out}}\rho _{ch}(\mathbf{x})d\mathbf{x}$ in $V_{out}$
is then given by%
\begin{equation}
G_{m}=\sum_{\mathbf{z}_{i}\in V_{out}}\frac{w(\mathbf{z}_{i})}{g}.
\label{Gm}
\end{equation}%
We note here that $G_{m}$ is a topological invariant under the map
deformation of $\hat{n}(x)$.

Let $\varepsilon $ be the scale of the core radius of monopoles, over which $%
\partial n$ varies. It follows from (\ref{rho}) that $\rho _{ch}(\mathbf{x}%
)\simeq (1/g)w(\mathbf{z}_{i})/\varepsilon ^{3}$. Let $w(z_{i})=w$ be equal
for all monopoles, the monopole density is then 
\begin{equation}
\rho _{m}(\mathbf{x})=\frac{\rho _{ch}(\mathbf{x})}{(2\pi /g)}\simeq \frac{w(%
\mathbf{x})}{2\pi \varepsilon ^{3}}  \label{density}
\end{equation}

Since the vacuum outside is colorless one must have $G_{m}=0$, which implies
that monopoles happened only in the pairs of monopole-anti-monopoles. The
length scale $\Lambda _{QCD}^{-1}$ of QCD can be introduced by QCD cutoff $%
\Lambda _{QCD}$. In the case $\Lambda _{QCD}=0.5GeV$, this scale is about $%
0.4fm$. When we choose $\varepsilon \simeq 0.4fm$ the monopole density then
mainly depends on $w(\mathbf{x})\,$, the winding numbers of $\hat{n}(\mathbf{%
x})$ at the local sites $\mathbf{x}$ of monopoles.

We now examine these multi-monopoles using the Skyrme-Faddeev(SF) model \cite%
{FN} as a magnetic dynamics. The SF model reads%
\begin{equation}
\mathfrak{L}^{SF}=\frac{\mu _{F}^{2}}{2}(\partial _{\mu }\hat{n})^{2}-\frac{%
\alpha }{4}(\partial _{\mu }\hat{n}\times \partial _{\nu }\hat{n})^{2}\text{,%
}  \label{SF}
\end{equation}%
The static energy is%
\begin{equation}
E^{SF}=\int d\mathbf{x}[\frac{\mu _{F}^{2}}{2}(\mathbf{\nabla }\hat{n})^{2}+%
\frac{\alpha }{4}(\partial _{i}\hat{n}\times \partial _{j}\hat{n})^{2}]
\label{HSF}
\end{equation}%
One takes, for simplicity, the $\hat{n}$-configuration to be ($%
n^{1}n^{2}n^{3}$) $=(\cos w\varphi \sin w\theta ,\sin w\varphi \sin w\theta
,\cos w\theta $), which has a winding of integer $w$. Direct calculation
shows that 
\begin{equation}
(\nabla \hat{n})^{2}\varpropto w^{2}.  \label{dn}
\end{equation}%
Owing to the topological reason, this proportionally also holds for an
alternative $w$-winding map $\hat{n}^{\prime }$ which is continuously
deformed from the above $\hat{n}$.

Using the virial theorem and (\ref{dn}), one can find that classical energy (%
\ref{HSF}) is 
\begin{equation}
E^{SF}=\int d\mathbf{x}\mu _{F}^{2}(\mathbf{\nabla }\hat{n})^{2}\varpropto
w^{2}.  \label{Ew}
\end{equation}%
We see that for a monopole with $w$-winding (i.e., magnetic charge $2\pi w/g$%
) its local energy (\ref{Ew}) is bigger that of a system of $w$ monopoles
with unit winding ($w=1$). Therefore, if the NP vacuum of QCD means it is
highly nontrivial in the sense that $\hat{n}(x)$ accommodate singularities
with nonzero winding densely distributed in this vacuum, such a vacuum can
be a stable system of monopolies with unit winding ($w=1$), in contrast with
the monopolies with higher winding ($\left\vert w\right\vert >1$).

For a bag with soft boundary, its boundary can be taken to be an across-over
region $V_{ao}$ between two vacua. Due to its complexity, we try to give a
rather qualitative picture for $V_{ao}$ in the viewpoint of dual
superconductor. Let us suppose that the variation of the monopole density $%
\rho _{m}(\mathbf{x})\varpropto w(\mathbf{x})$ by (\ref{density}) happens
mainly over $V_{ao}$. As $|\mathbf{x}|$ decreases, the topological
singularity decreases to vanishing over this region, which agrees with
analysis in section 2 that as $|\mathbf{x}|$ goes from inside of bag to
outside, the monopoles density increases from $0$ to a nonzero value $\rho
_{0}$. This is comparable with the core structure of the Abrikosov vortex in
type-II superconductor where the density of Cooper pairs rises from zero to
a uniform value as one goes from the core center to the outside of vortex.
In the region outside bag, the dominant variable is given by $\hat{n}$ and
the related energy is given by classical energy (\ref{HSF}).

\section{Abelian-Higgs phase and its model}

To obtain a calculational procedure with the dual superconductor mechanism,
we need a effective model for the across-over region $V_{ao}$. As discussed
in section 2, $\phi (x)$ in (\ref{dual}) can play the role of soliton field
interpolating in between the two vacua: $\phi (x)=0$ and $\phi (x)=v(\neq 0)$%
. It is then very useful to take the monopole density $\rho _{m}(\mathbf{x})$
to be proportional to the norm square of $\phi (x)$ in the (\ref{dual}): $%
\rho _{m}(\mathbf{x})\propto \left\vert \phi (x)\right\vert ^{2}$. In the
language of field theory, this implies we choose $\phi (x)$ as the monopole
field, similar to the wavefunction of Cooper pairs in the superconductor.
Writing $\phi (x)=\Phi (x)+\delta \phi $, where $\Phi (x)$ is the monopole
condensate and $\delta \phi $ its fluctuation, one has 
\begin{equation}
\langle \phi (x)\phi ^{\dag }(y)\rangle \approx \Phi (x)\Phi ^{\ast }(y),%
\text{for }x^{0}>y^{0}.  \label{ass3}
\end{equation}

In the bag picture of hadron with soft boundary region $V_{ao}$, there are
three scale regions: $V_{B}:=\{\mathbf{x|x}$ is in bag but not in $V_{ao}\}$%
, $V_{ao}$ and $V_{out}:=\{\mathbf{x|x}$ is outside of bag and $V_{ao}\}$,
in the increasing order of length scale.

As discussed in section 2, $V_{B}$ and $V_{out}$ can be taken to be in the
phase of the perturbative QCD phase and the NP condensate phase,
respectively. The relevant variables can be ultraviolet gluon field $A_{\mu
}^{a}$ ($a=1,2,3$) for the former and the infrared variable $\hat{n}$ for
the later. Here, we take ($A_{\mu }$,$\phi $) as the relevant variables for
the region $V_{ao}$, and derive the relevant model from (\ref{dual}) by
looking $\hat{n}$ as a background field. As will seen in the following, the
effective model for this region is the AH model, and we call the phase for
describing $V_{ao}$ the Abelian-Higgs phase.

Let us write 
\begin{equation}
\partial _{\mu }\hat{n}(x)=M\mathbf{e}_{\mu }(x)  \label{PM}
\end{equation}%
with $M=||\partial _{\mu }\hat{n}(\mathbf{x})||$. Clearly, $M\,\rightarrow 0$
when $\mathbf{x}\rightarrow 0$\ while $M\,\rightarrow M_{0}$ when $\mathbf{x}%
\rightarrow \infty $. For simplicity, we assume $M\simeq const<M_{0}.$ in $%
V_{ao}$. Then one can find $(\partial _{\mu }\hat{n})^{2}=M^{2}\{(\mathbf{e}%
_{0})^{2}-\sum_{i}(\mathbf{e}_{i})^{2}\}=-2M^{2}$, $H_{\mu \nu }=M^{2}h_{\mu
\nu }$ where $h_{\mu \nu }=\hat{n}\cdot (\mathbf{e}_{\mu }\times \mathbf{e}%
_{\nu })=\sin \theta _{\mu \nu }$, $\theta _{\mu \nu }$ is the angle between 
$\mathbf{e}_{\mu }$\textbf{\ }and\textbf{\ }$\mathbf{e}_{\nu }$ in the
iso-space. Also, 
\begin{eqnarray*}
\partial _{\mu }\hat{n}\cdot \partial _{\nu }\hat{n} &=&M^{2}\cos \theta
_{\mu \nu }, \\
n_{\mu \nu } &=&\eta _{\mu \nu }(\partial \hat{n})^{2}-\partial _{\mu }\hat{n%
}\cdot \partial _{\nu }\hat{n} \\
&=&-M^{2}(2\eta _{\mu \nu }-\cos \theta _{\mu \nu }). \\
(H_{\mu \nu })^{2} &=&M^{4}h_{\mu \nu }^{2}=\frac{M^{4}}{2}\sum_{\mu \nu
}(1-\cos 2\theta _{\mu \nu })
\end{eqnarray*}

For the magnetic field fluctuatation $H:=\langle (H_{\mu \nu })^{2}\rangle
^{1/2}$ one has%
\begin{eqnarray*}
H^{2} &=&\frac{M^{4}}{2}\left\{ \sum_{\mu \nu }\langle 1-\cos 2\theta _{\mu
\nu }\rangle \right\} \\
&\simeq &6M^{2}
\end{eqnarray*}%
where $\sum_{\mu \nu }1=12$. Here, we have used the random phase
approximation (RPA) 
\begin{equation*}
\sum_{\mu \nu }\langle \cos 2\theta _{\mu \nu }\rangle \simeq 0\text{.}
\end{equation*}%
Then, one has $M^{2}=H/\sqrt{6}$. The reformulated YM Lagrangian (\ref{dual}%
) then becomes%
\begin{eqnarray*}
\mathfrak{L}^{YM} &=&-\frac{1}{4}F_{\mu \nu }^{2}+\frac{M^{2}Z(\phi )}{4g}%
F^{\mu \nu }h_{\mu \nu }-\frac{M^{4}Z(\phi )^{2}}{4g^{2}}h_{\mu \nu }^{2} \\
&&+\frac{M^{2}}{2g^{2}}\{[2\eta _{\mu \nu }+\cos \theta _{\mu \nu }+i\sin
\theta _{\mu \nu }](\nabla ^{\mu }\phi )^{\dag }\nabla ^{\nu }\phi +h.c\},
\end{eqnarray*}

In the RPA, one has $\left\langle h_{\mu \nu }\right\rangle \simeq 0$, $%
\left\langle h_{\mu \nu }^{2}\right\rangle \simeq 6$, $\langle e^{i\theta
_{\mu \nu }}(\nabla ^{\mu }\phi )^{\dag }\nabla ^{\nu }\phi \rangle \simeq 0$%
. Then, one has the following averaged Lagrangian 
\begin{equation}
\mathfrak{L}^{AH}=-\frac{1}{4}F_{\mu \nu }^{2}+\frac{2H}{\sqrt{6}g^{2}}%
(\nabla ^{\mu }\phi )^{\dag }\nabla ^{\nu }\phi -\frac{H^{2}}{4g^{2}}\langle
Z(\phi )^{2}\rangle  \label{LAH}
\end{equation}%
where (\ref{ass3}) is used:$\langle (\nabla ^{\mu }\phi )^{\dag }\nabla
_{\mu }\phi \rangle =\left( \nabla _{\mu }\Phi (x)\right) ^{\ast }\nabla
^{\mu }\Phi (x)$. \ 

Using the Wick theorem and the Bose symmetry of the scalar field, one finds 
\begin{eqnarray*}
\left\langle (\phi ^{\dag }\phi )^{2}\right\rangle &=&\left\langle \phi
^{\dag }\phi \right\rangle \left\langle \phi ^{\dag }\phi \right\rangle
+\left\langle \phi ^{\dag }\phi ^{\dag }\right\rangle \left\langle \phi \phi
\right\rangle +\left\langle \phi ^{\dag }\phi \right\rangle \left\langle
\phi ^{\dag }\phi \right\rangle \\
&=&2\left\langle \phi ^{\dag }\phi \right\rangle ^{2}
\end{eqnarray*}%
\begin{eqnarray*}
\left\langle Z(\phi )^{2}\right\rangle &=&\left\langle 1+(\phi ^{\dag }\phi
)^{2}-2\phi ^{\dag }\phi \right\rangle \\
&\approx &1+2|\Phi ^{\ast }\Phi |^{2}-2\Phi ^{\ast }\Phi \\
&=&2(|\Phi |^{2}-1/2)^{2}+1/4.
\end{eqnarray*}%
Using above relations and rescaling $\Phi $ to that with dimension of mass 
\begin{equation}
\sqrt{\frac{3}{2}}\frac{m}{g}\Phi (x)\rightarrow \Phi (x),  \label{rep}
\end{equation}%
we arrive at the following dual AH model with a constant added 
\begin{equation}
\mathfrak{L}^{eff}=-\frac{1}{4}F_{\mu \nu }^{2}+|(\partial _{\mu }-igA_{\mu
})\Phi |^{2}-V(\Phi )-\frac{H^{2}}{8g^{2}}.  \label{AHM}
\end{equation}%
where the replacement (\ref{rep}) was used. The potential $V(\Phi )$ is
given by 
\begin{equation}
V(\Phi )=\frac{\lambda ^{2}}{4}(|\Phi |^{2}-v^{2})^{2},  \label{Vphi}
\end{equation}%
\ where 
\begin{eqnarray}
\lambda &=&\sqrt{3}g,  \notag \\
v &=&\frac{\sqrt{H}}{\sqrt[4]{6}g}  \label{m1}
\end{eqnarray}%
It has the Mexico-hat form, implying two vacua $\Phi =0$ and $\Phi =v$. As
mentioned before, $\Phi $ is assumed, up to a constant, to be the monopole
condensate. So, the\ two vacua correspond to the perturbative vacuum and NP
vacuum, as expected in section 3. The static energy associated with the dual
AH model (\ref{AHM}) can be given by%
\begin{equation}
E^{AH}=\int_{V_{ao}}d\mathbf{x\{}\frac{1}{2}\vec{B}^{2}+|D_{i}\Phi
|^{2}+V(\Phi )+\frac{H^{2}}{8g^{2}}\}.  \label{DAH}
\end{equation}

\section{Glueball energy}

The model (\ref{AHM}) is nothing but the dual AH model suggested as the
effective model of the dual superconductor picture \cite{Suzuki} for the
confining phase of QCD. It is known that this model admits the
Nielsen-Olesen vortex solution \cite{Nielsen-Olesen} and the dual Meissner
effect is measured by two scales: the coherent length $\xi =1/m_{\Phi }$ and
penetrating length $\lambda _{L}=1/m_{A}$. For the studies on the
Abelian-Higgs model as a long-distance gluodynamics in the lattice
framework, see \cite{AHM}.

The masses $m_{\Phi }$ for the Higgs field $\Phi $ and $m_{A}$ for the
chromo-electric field $A_{\mu }$ can be determined by the potential
parameter $\lambda $ and $v$ in (\ref{m1}). They are 
\begin{eqnarray}
m_{\Phi } &=&\sqrt{\lambda }v=\frac{\sqrt{3H}}{\sqrt[4]{6}},  \notag \\
m_{A} &=&\sqrt{2}gv=\frac{\sqrt{2H}}{\sqrt[4]{6}}\text{ }  \label{mA}
\end{eqnarray}%
With (\ref{mA}), one finds that the Ginzburg-Landau parameter for the NP
vacuum medium is 
\begin{equation}
\kappa =\frac{m_{\Phi }}{m_{A}}=\frac{\sqrt{3}}{\sqrt{2}}\text{, (type-II).}
\label{para}
\end{equation}%
The result (\ref{para}) predicts the vacuum type of type-II superconductor.
The Nielsen-Olesen vortex solution indicates that $\Phi $ increases from
zero near the vortex core and approaches a nonzero constant $v$ far away
from the vortex core.

When the stable gluon flux confined in bag, one can expect that the energy
for gluon field within bag stabilizes the normal vacuum $\Phi =0$ by
compensating energy density 
\begin{equation}
\frac{H^{2}}{8g^{2}}=V(0)-V(v).  \label{DD}
\end{equation}%
Here, the bag is taken to be wall limit of confined multi-vortices \cite%
{BolognesiNum06}.

In the cylindrically symmetric case the field strength in $V_{ao}$ is
written as $B=\nabla \times A(r)$, where $A(r)$ is the nonvanishing
azimuthal component of $A_{i}\allowbreak $, and the gluon field in $V_{B}$
as $(B,B,B)$. The gluon energy in $V_{B}$ is given by $E_{A}=(3B^{2}/2)V_{B}$%
. Collecting the energies in all regions one has%
\begin{equation}
E=\frac{3B^{2}}{2}V_{B}+E^{AH}+E^{SF}  \label{E}
\end{equation}%
Here, $E^{SF}$ in (\ref{E}) is taken to be the energy in $V_{out}$. Owing to
the requirement of continuety and approximated uniformity of condensate in $%
V_{out}$, the energy density $u_{0}$ in SF model equals approximately the
dual AH energy density at the boundary of $V_{ao}$ and $V_{out}$: $%
u_{0}\approx H^{2}/(8g^{2})$. One then gets 
\begin{eqnarray}
E &=&\frac{3B^{2}}{2}V_{B}+\frac{B^{2}}{2}V_{ao}+\int_{V_{ao}+V_{out}}d%
\mathbf{x}\frac{H^{2}}{8g^{2}}  \notag \\
&&+\int_{V_{ao}}d\mathbf{x\{}|D_{i}\Phi |^{2}+V(\Phi )\}.  \label{GluE}
\end{eqnarray}

Let $R$ be the size of bag, $C$ the bag equator with the section $A(C)$.
Being a vortex formed in normal vacuum ($\Phi =0$), the chromo-electric flux
passing through $A(C)$ is quantized by the monopole condensate field $\Phi
=\rho \exp (iN\theta )$ with $N$-multiply quantized vortices $\Phi
_{A}(C)=2N\pi /g$, with $N$ being the quanta number of the vortex within the
bag. Notice that $\Phi _{A}(C)\approx B\pi R^{2}$ and thereby $B=2N/(gR^{2})$%
. Adding the energy [$V(0)-V(\Phi _{0})]V_{B+ao}$, which is due to the
vacuum energy density difference (\ref{DD}), and throwing away the infinite
constant integration over $V_{ao}+V_{out}$, we obtain the glueball energy 
\begin{eqnarray}
E &=&\frac{2N^{2}}{g^{2}R^{4}}[2V_{B}+V_{B+ao}]+\frac{H^{2}}{8g^{2}}%
V_{B+ao}+\int_{V_{ao}}d\mathbf{x\{}|D_{i}\Phi |^{2}+V(\Phi )\}  \notag \\
&=&\frac{8\pi N^{2}}{3g^{2}R}\left[ 1+2(1-\frac{\lambda _{L}}{R})^{3}\right]
+\frac{\pi H^{2}}{6g^{2}}R^{3}  \notag \\
&&+\int_{V_{ao}}d\mathbf{x\{}|D_{i}\Phi |^{2}+V(\Phi )\}  \label{Glu}
\end{eqnarray}%
where we choose $V_{B+ao}:=V_{B}+V_{ao}=\frac{4}{3}\pi R^{3}$, $V_{B}=4\pi
(R-\lambda _{L})^{3}/3$ and $V_{ao}=4\pi \lbrack R^{3}-(R-\lambda
_{L})^{3}]/3$. Here, the bag boundary thickness was chosen to be $\lambda
_{L}$, which equals approximately $1/m_{A}=\sqrt[4]{6}/\sqrt{2H}$. The bag
wall tension can be given by $T_{W}:=(1/V_{ao})\int_{V_{ao}}d\mathbf{x\{}%
|D_{i}\Phi |^{2}+V(\Phi )\}$. We see that the first two terms have the form
of MIT-bag energy in the thin-wall limit $\lambda _{L}/R\rightarrow 0$.
Minimization of the energy (\ref{Glu}) fixes $R$ as a function of ($N,g$,$H$%
). Recall that $m_{\Phi }^{2}\propto H\propto \langle (\nabla \hat{n}%
)^{2}\rangle $ (see the relation (\ref{mA}), one knows that the dual AH
model (\ref{AHM}) and (\ref{Glu}) provide us with a calculational procedure
for the glueball energy, with two parameters $H$ and $N$. Here, $g$ can be
chosen as $g_{s}=(4\pi \alpha _{s})^{1/2}$.

We note here that our framework for computing glueball energy is comparable
with that of the holographic dual model \cite{Boschi-Filho,Teramond05} of
QCD based on AdS/QCD correspondence \cite{Polchinski02}. This can be seen
from the following remarks on the two frameworks: (1) in modeling the
glueballs both employ the "string/field" correspondence or "duality" . Ours
is of the "electric-magnetic" duality, which has a gravitational analogy
with black hole in color space \cite{Jiadj07}; The holographic model is
based on the supergravity duality of QCD \cite{Witten}; (2) both introduce a
finite cutoff to truncate the regime where conformal field modes (the
massless gluon field modes for the former and the string modes for the
later) can propagate. (3) In the "hard-wall" or "thin-wall" limit both
provide an analogy of the MIT bag model. The bag is described by step
function given by the scalar condensate $\Phi $ in our framework and by a
metric factor in holographic model. In spite of these similarities, one can
see that our model differs from the holographic model (e.g., the AdS slice
dual model \cite{Boschi-Filho}) in that the field modes being confined in
bag in our model are the flux tubes of gluon field in the form of
multi-vortices while the counterparts in the holographic model are the
lightest string modes in high dimensional string theory \cite{Teramond05}.
Therefore, the duality in our model is actually that between field and
vortices which end on the bag boundary, and can be viewed as the prototype
of the string/field duality in string theory within the framework of field
theory.

Explicit calculation of the glueball mass depends on the solution to the
dual AH model (\ref{AHM}) which is to be used to calculate the last
integration concerning the bag wall tension $T_{W}$ in (\ref{Glu}). The
magnetic condensation $H$ can be given by the one-loop effective potential
calculation \cite{Cho02} $\sqrt{H}=\Lambda \exp (-\frac{6\pi ^{2}}{%
11g_{s}^{2}}+\frac{1}{2})$ , where $\Lambda $ is the QCD cutoff ($\simeq
0.3\sim 0.5GeV$). The further calculations and the comparison with the
lattice prediction $M_{0^{++}}=1.61\pm 0.15GeV$ \cite{Teper}\ as well as
holographic prediction $1.3GeV$ (for $\Lambda =0.26GeV$) for the mass of
glueball $0^{++}$ will be presented in the forthcoming paper.

\section{Summary}

The dual structure of the SU(2) YM theory is revisited associated with the
bag picture of hadron and using the reparametrization called connection
decomposition. It is shown that theory admits an Abelian-Higgs phase
effectively described a dual Abelian-Higgs model, with a Higgs vacuum
constant added. This phase corresponds to the soft boundary region of the
bag which is the across-over region between the normal vacuum and NP vacuum
of QCD. Applying the bag picture for glueball, we presented a calculation
procedure for glueball energy based on the idea of wall-vortices.

D. Jia is grateful to C-R Ji for his hospitality and valuable discussions
during author's visit to NC State University; Author also thanks C. Liu, P.
Wang for discussions.


\begin{thebibliography}{99}
\bibitem{BolognesiNum06} Bologonesi S, Nucl. Phys. B 2005, \textbf{730}
:127. [arXiv:hep-th/0507273]; Nucl. Phys. B 2006, \textbf{741} :1
[arXiv:hep-th/0512132].

\bibitem{Jiadj07} Jia D, Ai D Z, HEP \& NP, 2007, 31(5):64 . (in Chinese) ()
[arXiv:hep-th/0605136];

\bibitem{Duan} Duan Y S, Ge M L, Sci. Sin. 1979 \textbf{11} :1072. (in
Chinese).

\bibitem{Cho80} Cho Y M, Phys. Rev. D 1980, \textbf{21}:1080.

\bibitem{FN} Faddeev L D, Niemi A J, Phys. Rev. Lett. 1999, \textbf{82}
:1624.

\bibitem{Langmann} Langmann E, Niemi A J, Phys. Lett. B 1999, \textbf{463}:
252.

\bibitem{tHooftA03} 't Hooft G, Nucl. Phys. A 2003 \textbf{721} :30
[arXiv:hep-th/0207179].

\bibitem{TDLee} Lee T D, \emph{Particle Physics and Introduction to Field
Theory}, (Harwood Academic, Amsterdam, 1983).

\bibitem{WIL89} Wilets L, \emph{Nontopological Soliton}, World Scientific
Lecture Notes in Physics, Vol. 24, (World Scientific, Singapore, 1989).

\bibitem{Jinstant} Jia D, Duan Y S, Mod. Phys. lett. A 2001, \textbf{16}
:1863; Duan Y S et al. J. Math. Phys. 2000, \textbf{41} :4379.

\bibitem{JDJ} Jia D et al. HEP \& NP. 2003, \textbf{4} :293.

\bibitem{Gribov} Gribov V N, Orsay lectures on confinement (II), in \emph{%
The Gribov Theory of Guark Confinement}, Ed. J. Nyiri, (World Scientific,
Singapore, 2001).

\bibitem{Nambu} Numbu Y, Phys. Rev. D. 1974, \textbf{10} :4262; 't Hooft G,
in High Energy Physics, edited by A. Zichichi, EPS International Conference,
Palermo,1975 (Editrice Compositori, Bologna, 1975); Mandelstam S, Phys. Rep.
C 1976, \textbf{23} :245.

\bibitem{tHooftB455} 't Hooft G, Nucl. Phys. B 1981, [\textbf{FS3}] \textbf{%
190} :455.

\bibitem{AHM} Kato S, et al., Nucl. Phys. B 1998, \textbf{520} :323;
Schilling K, Bali G S et al. Nucl. Phys. 1998 (Proc. Suppl.) 63:519; Gubarev
F V et al. Phys. Lett. B 1999, 468 : 134.

\bibitem{Suzuki} Suzuki T, Prog. Theor. Phys. 1988, 80 :929.

\bibitem{Nielsen-Olesen} Nielsen H B, Olesen P, Nucl. Phys. B 1973, 61:45.

\bibitem{Boschi-Filho} Boschi-Filho H, Braga N R\ F, Eur. Phys. J. C 192004,
32:529; J. High Energy Phys. 2003, 05: 009.

\bibitem{Teramond05} Teramond Guy F de, Brodsky S J, Phys. Rev. Lett.\ 2005,
94 :201601; 

\bibitem{Polchinski02} Polchinski J, Strassler M J, Phys. Rev. Lett.\ 2002,
88 :031601.\ 

\bibitem{Witten} Witten E, Adv. Theor. Math. Phys. 1998, 2 :5051. Gross D J,
Ooguri H, Phys. Rev. D 1998, 58 :106002.

\bibitem{Cho02} Cho Y M, Pak D G, Phys. Rev. D 2002, 65 :074027.\ 

\bibitem{Teper} Teper M J, hep-lat/9711011
\end{thebibliography}
\end{document}